\documentclass[aps,prl,twocolumn,groupedaddress,longbibliography]{revtex4-1}

\usepackage{amsmath}
\usepackage{amsthm}
\usepackage{mathrsfs}
\usepackage{hyperref}
\usepackage{txfonts}
\usepackage{color}
\usepackage{graphicx}
\usepackage{tikz}

\newtheorem{lemma}{Lemma}
\newtheorem*{lemma*}{Lemma}

\newcommand{\PP}{\mathcal{P}^{(\text{1g})}}   
\newcommand{\CC}{\mathcal{C}^{(l_1)}}             
\newcommand{\PPbar}{\bar{\mathcal{P}}^{(\text{1g})}}   
\newcommand{\CCbar}{\bar{\mathcal{C}}^{(l_1)}}             

\begin{document}


\title{Wave-Particle-Mixedness Complementarity}

\author{Xiangdong Zhang$^{1,2,4}$}
\author{Jiahao Huang$^{1}$}
\author{Min Zhuang$^{1}$}
\author{Xizhou Qin$^{1}$}

\author{Chaohong Lee$^{1,3}$}
\altaffiliation{Corresponding author. Emails: lichaoh2@mail.sysu.edu.cn; chleecn@gmail.com}

\affiliation{$^{1}$TianQin Research Center \& School of Physics and Astronomy, Sun Yat-Sen University (Zhuhai Campus), Zhuhai 519082, China}

\affiliation{$^{2}$School of Physics, Sun Yat-Sen University (Guangzhou Campus), Guangzhou 510275, China}

\affiliation{$^{3}$State Key Laboratory of Optoelectronic Materials and Technologies, Sun Yat-Sen University (Guangzhou Campus), Guangzhou 510275, China}

\affiliation{$^{4}$Beijing Computational Science Research Center, Beijing 100084, China}

\date{\today}

\begin{abstract}
  Wave-particle duality, an important and fundamental concept established upon pure quantum systems, is central to the complementarity principle in quantum mechanics.
  However, due to the environment effects or even the entanglement between the quanton and the which-way detector (WWD), the quanton should be described by a mixed quantum state but not a pure quantum state.
  Although there are some attempts to clarify the complementarity principle for mixed quantum systems, it is still unclear how the mixedness affects the complementary relation.
  Here, we give a ternary complementary relation (TCR) among wave, particle and mixedness aspects for arbitrary multi-state systems, which are respectively quantified by the $l_1$ measure for quantum coherence, the which-path predictability, and the linear entropy.
  In particular, we show how a WWD can transform entropy into predictability and coherence.
  Through modifying the POVM (positive-operator valued measure) measurement on WWD, our TCR can be simplified as the wave-mixedness and particle-mixedness duality relations.
  Beyond enclosing the wave-particle duality relation [PRL \textbf{116}, 160406 (2016)], our TCR relates to the wave-particle-entanglement complementarity relation [PRL \textbf{98}, 250501 (2008); Opt. Commun. \textbf{283}, 827 (2010)].
\end{abstract}


\maketitle

%
%
%
%
\emph{Introduction}.---Quantum objects may possess complementary properties that are equally real but mutually exclusive.
This is descirbed by the famous complementarity principle~\cite{Bohr28}, that is, although the complementary properties can be observed separately, but they can never exhibit fully at the same time.
This principle can be demonstrated by the wave-particle duality in quantum interferometry, where the particle property is understood as ``the quanton (an abbreviation of quantum object that go through the interferometer) has a definite path'' and the wave property is understood as ``the quanton can interfere with itself''.
In one measurement, we can precisely observe either the quanton's path or interference pattern, but not both.
The wave-particle duality in a two-path interferometry is ofen descirbed by the Englert-Greenberger-Yasin (EGY) inequality~\cite{Greenberger88, Jaeger95, Englert96, Englert_erasure},
\begin{equation}
\label{eq:EGYieq}
\mathcal{D}^2+\mathcal{V}^2 \leq 1.
\end{equation}
The distinguishability $\mathcal{D}$ describes the particle property, while the visibility $\mathcal{V}$ quantifys the wave property.
The EGY inequality has been experimentally verified via various systems, such as, single atoms~\cite{Durr98}, faint lasers~\cite{Schwindt99}, nuclear spins~\cite{Peng03, Peng05} and single photons~\cite{Jacques08}.
In addition to the conventional wave-particle duality, the complementarity principle has been understood in views of information which is expressed as the entropic uncertainty~\cite{Wootters79, Coles14, Coles16, Coles17}.

Beyond two-state systems, the EGY inequality has been generalized to multi-state systems~\cite{Jaeger95, Coles16, Englert08, Durr01, Bimonte03, Siddiqui15, Bera15, Qureshi16, Bagan16}.
Given the $l_1$ measure for quantum coherence $\CC$~\cite{Baumgratz14} and the which-path predictability $P_s$, a successful generalization of the EGY inequality~(\ref{eq:EGYieq}) reads as~\cite{Bagan16},
\begin{equation}
\label{eq:EGYieq_multi}
\left( \frac{N}{N-1}P_s - \frac{1}{N-1} \right)^2+\left( \CC \right)^2 \leq 1,
\end{equation}
where $N$ is the number of states.
Although this inequality is valid for both pure and mixed quantum systems, it does not explore how the mixedness aspect affects the complementarity principle.
In particular, it is important to clarify whether one can transform mixedness aspect into wave or particle aspect?

In this Letter, we present a wave-particle-mixedness complementarity relation, which is expressed as a ternary complementary relation (TCR).
In addition to the $l_1$ measure for quantum coherence and the which-path predictability, which respectively quantify the wave and particle aspects, our TCR includes the linear entropy as the third term, which quantifies the mixedness aspect.
Naturally, as the linear entropy is always a non-negative number, our wave-particle-mixedness TCR always ensures the validity of the duality relation between coherence and predictability~\cite{Bagan16}.
Moreover, by adjusting the POVM (positive-operator valued measure) measurement on WWD (which-way detector), we show how the mixedness apsect can be transformed into the wave and particle aspects.
Furthermore, as the mixedness aspect of a single-partite reduced density matrix relates to the corresponding bi-partite entanglement, our TCR relates to the complementarity relation among single-partite coherence, single-partite predictability and bi-partite entanglement~\cite{Oppenheim03,Peng05,Melo07,Peng08,Jakob10,Fedrizzi11}.
In particular, when the which-path predictability is fixed under suitable POVM measurements on WWD, our TCR gives a wave-mixedness duality relation, which is reminiscent of the complementarity between entanglement and coherence~\cite{Singh15,Streltsov16,Chitambar16}.

\emph{TCR for Multi-State Quanton without WWD}.---We first consider the wave-particle-mixedness complementarity of an arbitray $N$-state quantum system without WWD.
It can be illustrated by an $N$-path interferometry: a quanton goes through the first beam splitter, accumulates relative phases and then goes through the second beam splitter for interference.
The which-path predictability between two beam splitters and the interference visibility outside the second beam splitter are usually used to quantify the particle and wave aspects of the quanton.
For an arbitray $N$-state quantum system, one may choose an observable $\hat{O}$ which specifies a set of eigenstates $\{ \lvert i \rangle \}$.
Therefore, the quanton's state can be described by an $N \times N$ density matrix $\rho$ under the basis $\{ \lvert i \rangle \}$.
The which-path predictability is defined as the so called ``one guess bet'' measure~\cite{Jaeger95, Bimonte03, Englert08},
\begin{equation}
\PP \left( \rho \right) \equiv \frac{N}{N-1} p_1 - \frac{1}{N-1}, \quad p_1 \equiv \max_k \rho _{kk}.
\label{eq:def_P1g}
\end{equation}
Here, $\PP$ is the normalized success probability for guessing the measurement result of $\hat{O}$ before actually measuring it.
There are several different quantities for characterizing the wave aspect, we use the $l_1$ measure for quantum coherence~\cite{Baumgratz14},
\begin{equation}
\label{eq:def_Cl1}
\CC \left( \rho \right) \equiv \frac{1}{N-1} \sum_{i \neq k} \lvert \rho _{ik} \rvert.
\end{equation}
This quantity represents the quanton's ability to produce interference pattern when the relative phases between different paths (described by the basis $\hat{O}$) are changed.

Similar to the inequality~(\ref{eq:EGYieq_multi}), there exists a duality relation between the predictability $\PP$ and the coherence $\CC$,
\begin{equation}
\label{eq:PCduality}
\left(  \PP  \right) ^2 + \left(  \CC  \right) ^2 \leq 1.
\end{equation}
However, the above inequality can only be saturated by some pure states and the equal sign can never hold for mixed states.
Can one find a new complementary relation whose equal sign can even hold for mixed states?
By introducing the normalized linear entropy, $S_L \left(\rho\right) = \left.N(1-\mathrm{tr}\{\rho^2\})/(N-1)\right.$, we find a TCR,
\begin{equation}
\label{eq:PCStriality}
\left(\PP\right) ^2 + \left(\CC\right) ^2 + S_L \leq 1,
\end{equation}
which is valid for arbitrary $N \times N$ density matrices and become an equality when $N=2$.
Since $S_L$ is always non-negative, the TCR~(\ref{eq:PCStriality}) naturally enclose the duality relation~(\ref{eq:PCduality}).
To prove the above TCR~(\ref{eq:PCStriality}), we introduce the following two lemmas:
\begin{lemma}
\label{lemma:diag}
For an arbitrary $N \times N$ density matrix $\rho$, we always have
\begin{equation}
\label{eq:lemmaDiag}
\sum_{k=1}^{N} \left( \rho _{kk} \right) ^2 \geq p_1^2 + \frac{\left( 1 - p_1 \right) ^2}{N-1}
\end{equation}
with $p_1 = \max_k \rho _{kk}$ being the largest diagonal element of $\rho$.
The equal sign holds if and only if all diagonal elements of $\rho$, except the maximal one, have the same value.
\end{lemma}
\begin{lemma}
\label{lemma:offdiag}
For an arbitrary $N \times N$ density matrix $\rho$, we always have
\begin{equation}
\label{eq:lemmaOffdiag}
\sum _{i \neq k} \sum _{l \neq m} \lvert \rho _{ik} \rvert \cdot \lvert \rho _{lm} \rvert \leq N(N-1)\sum _{i \neq k} \lvert \rho _{ik} \rvert ^2.
\end{equation}
The equal sign holds if and only if all off-diagonal elements of $\rho$ have the same modulus.
\end{lemma}

The proof of these two lemmas is just some technical jobs (see the Supplemental Material~\ref{app:lemmaProof} for details).
Now we show how to prove the TCR~(\ref{eq:PCStriality}) by using these two lemmas.
Given a density matrix $\rho$, the left-hand side (LHS) of Eq.~(\ref{eq:PCStriality}) reads as
\begin{subequations}
\begin{eqnarray}
\nonumber
& &\left( N-1 \right) ^2 \left[ %
                     \textcolor{blue}{\left( \PP \right) ^2} + %
                     \textcolor{red}{\left( \CC \right) ^2} + %
                     S_L \right]
\\
\nonumber
&=&\textcolor{blue}{\left( N p_1 -1 \right) ^2} + %
      \textcolor{red}{\left( \sum _{i \neq k} \lvert \rho _{ik} \rvert \right) ^2} + %
      N (N-1) \left( 1 - \sum _{i, k} \lvert \rho _{ik} \rvert ^2 \right)
\\
\label{eq:PCStriProof_l1}
&=&\textcolor{blue}{\left( N p_1 -1 \right) ^2}             +          (N-1)                -            N (N-1) \sum _{k} \left( \rho _{kk} \right) ^2
\\*
\label{eq:PCStriProof_l2}
& &+\textcolor{red}{\left( \sum _{i \neq k} \lvert \rho _{ik} \rvert \right) ^2}            -             N (N-1) \sum _{i \neq k} \lvert \rho _{ik} \rvert ^2
\\*
\nonumber
& &+\left( N-1 \right) ^2.
\end{eqnarray}
\end{subequations}
By using Lemma~\ref{lemma:diag}, the term Eq.~(\ref{eq:PCStriProof_l1}) becomes zero if $\sum_k \left(\rho_{kk} \right)^2$ is taken its lower bound, this means that the term Eq.~(\ref{eq:PCStriProof_l1}) is always non-positive.
By using Lemma~\ref{lemma:offdiag}, as $\left(\sum_{i \neq k} \lvert \rho_{ik} \rvert \right)^2 = \sum_{i \neq k} \sum_{l \neq m} \lvert \rho_{ik} \rvert \cdot \lvert \rho_{lm} \rvert$, we have the term Eq.~(\ref{eq:PCStriProof_l2}) is always non-positive.
Therefore, from the non-positive property of Eq.~(\ref{eq:PCStriProof_l1}) and Eq.~(\ref{eq:PCStriProof_l2}), we have
\begin{equation}
\label{eq:PCStriality2}
\left(N-1 \right)^2 \left[\left(\PP\right)^2 + \left(\CC\right)^2 + S_L\right] \leq \left(N-1\right)^2,
\end{equation}
which is equivalent to the TCR~(\ref{eq:PCStriality}).
Since both Lemma~\ref{lemma:diag} and Lemma~\ref{lemma:offdiag} are used in the proof, the equal sign of our TCR~(\ref{eq:PCStriality2}) and (\ref{eq:PCStriality}) hold if and only if the equal sign in both Lemma~\ref{lemma:diag} and Lemma~\ref{lemma:offdiag} hold together.
These conditions request that the density matrix $\rho$ should be in the form of
\begin{equation}
\label{eq:eqsignHold}
\rho = \begin{pmatrix}
p_1 &          &         &             \\
       & p_2   &         & \lvert a \rvert \text{e}^{-\iota \varphi _{ik}}      \\
\lvert a \rvert \text{e}^{\iota \varphi _{ik}}    &      &       \ddots  &        \\
     &      &     & p_2
\end{pmatrix}.
\end{equation}
Here, $\iota = \sqrt {-1}$ is the imaginary unit, $\lbrace \varphi_{ik} \vert i > k \rbrace$ are arbitrary phases, all off-diagonal elements have the same modulus $\lvert a \rvert$, and $p_1$ is the maximal diagonal element while $p_2 = \left. (1 - p_1)/(N - 1) \right.$ is the remaining diagonal elements.
More generally, $p_1$ does not necessarily be $\rho_{11}$, it can also be other diagonal element $\rho_{22}$ or $\rho_{33}$ etc.
Since a density matrix should be positive-semidefinite, we have $\lvert a \rvert \leq \sqrt {p_1 p_2}$ for $N=2$ and $\lvert a \rvert \leq p_2$ for $N>2$.
For a two-state quanton ($N=2$), the density matrix~(\ref{eq:eqsignHold}) actually includes all possible $2\times2$ density matrix and the corresponding TCR~(\ref{eq:PCStriality}) becomes an equality.

\paragraph{Remark 1.}
The linear entropy $S_L$, a quantitative description of the degree of mixedness, can be viewed as a concept opposite to the purity $P=tr\{\rho^2\}$.
That is, larger purity means less linear entropy, and vice versa.
We always have $\left. S_L = 0 \right.$ for a pure state and $\left. S_L = 1 \right.$ for a maximally mixed state $\left. \rho = \text{diag} \{ 1/N, 1/N, \dots \} \right.$.
A mixed state can be understood as a classical summation of several pure quantum states and so that $S_L$ also measures how ``classical'' the quanton is.

\paragraph{Remark 2.}
The linear entropy $S_L\left( \rho \right)$ is a concave function of the density matrix $\rho$, that is,
\begin{equation}
\label{eq:concaveRelation}
\begin{split}
S_L\left( a \rho_1 + b \rho_2 \right) \geq  a S_L \left( \rho_1 \right) + b S_L \left( \rho_2 \right),
\\
\textrm{with}~a + b = 1~\textrm{and}~a \in \left[ 0, 1 \right].
\end{split}
\end{equation}
This is a bit different from $\PP$ and $\CC$, which are convex.
The proof for the above inequality is given in the Supplemental Material~\ref{app:concaveProof}.

\paragraph{Remark 3.}
The inequality~(\ref{eq:PCStriality}) describes a ternary complementarity relation among which-path predictability, quantum coherence and linear entropy.
The which-path predictability $\PP$ is the probability that we correctly guess the measured value of the considered observable $\hat{O}$, which quantitatively measures ``particle'' aspect.
The quantum coherence $\CC$ measures how good a system can be in a quantum superposition state and keep stable relative phases between different paths labeled by the eigenstates of the considered observable $\hat{O}$.
Although a little abstract, quantum coherence is a quantitative measure of ``wave'' aspect.
Finally, the linear entropy $S_L$ is a measure of the degree of mixedness.
The TCR~(\ref{eq:PCStriality}) indicates that the total amount of wave-particle-mixedness aspects is bounded.
For an example, if a quanton is a classical object (i.e. a fully mixed object), it neither looks like a particle nor a wave since $\PP = \CC = 0$ regardless of the choice of $\hat{O}$.

\paragraph{Remark 4.}
Based upon D\"{u}rr's definition for predictability and visibility~\cite{Durr01}, one can easily obtain the following three-term equality,
\begin{equation*}
\left(\mathcal{P}^{(\text{D\"{u}rr})}\right)^2 + \left(\mathcal{V}^{(\text{D\"{u}rr})}\right)^2 + S_L = 1,
\end{equation*}
with
\begin{eqnarray*}
\mathcal{P}^{(\text{D\"{u}rr})} (\rho) &\equiv& \sqrt{ \frac{N}{N-1} \sum_j \left(\rho_{jj} - \frac 1N \right)^2},
\\
\mathcal{V}^{(\text{D\"{u}rr})} (\rho) &\equiv& \sqrt{ \frac{N}{N-1} \sum_{j \neq k} \lvert \rho _{jk} \rvert^2}.
\end{eqnarray*}
However, this three-term equality seems a bit trivial since one can obtain it by just using the basic property of $\text{tr} \bigl\{ \rho^2 \bigr\}$.

\begin{figure}[ht]
\includegraphics[width=0.5\textwidth]{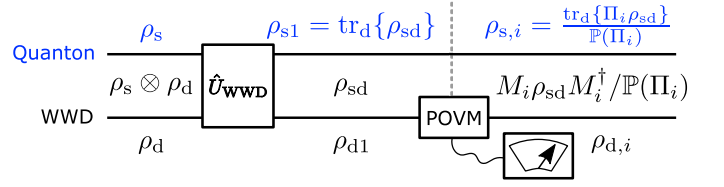}%
\caption{\label{fig:quantonWWDinteract}
  Schematic diagram for how the POVM on WWD affects the quanton. The input state is $\rho_\text{s} \otimes \rho_\text{d}$ with $\rho_\text{s}$ and $\rho_\text{d}$ respectively denoting the initial states of the quanton and the WWD. After the interaction $\hat{U}_\text{WWD}$, the quanton and the WWD are respectively described by the reduced density matrices $\rho_\text{s1}$ and $\rho_\text{d1}$. The set of POVM on WWD is described by $\{\Pi_i = M_i^\dag M_i\}$ with a set of measurement operators $\{M_i\}$. $\rho_{\text{s},i}$ and $\rho_{\text{d},i}$ are the reduced density matrices of quanton and WWD for the $i$-th subensemble.
  }
\end{figure}

\emph{TCR for Multi-State Quanton with WWD}.---In above, we give the complementary relation~(\ref{eq:PCStriality}) by using \textit{a priori} knowledge: the density matrix.
Below we will discuss the case where the knowledge from measurement is used.
As shown in Fig.~\ref{fig:quantonWWDinteract}, we introduce a WWD which interacts with the quanton and then perform a POVM measurement on the WWD.
Therefore, the final information of the quanton will depend on the POVM.
\begin{figure*}
\includegraphics[width=\textwidth]{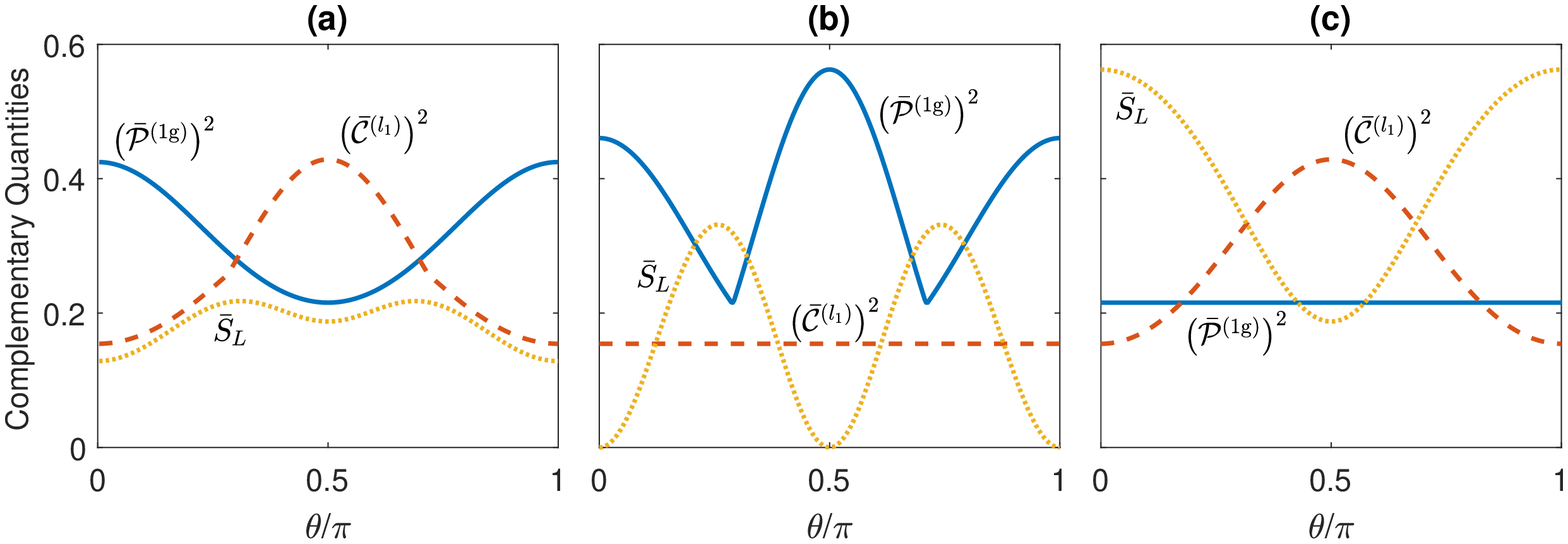}
\caption{\label{fig:POVMsweep}
  Transformation among quantum coherence, which-path predictability and linear entropy for a three-state quanton under POVM measurements on WWD.
  The quanton's initial state is $\rho_\text{s}=\lvert \psi_\text{s}\rangle \langle \psi_\text{s} \rvert$ with $\langle \psi_\text{s} \rvert \propto \left( 1, 3, 2 \right)$.
  The WWD's states are $U_i \rho_\text{d} U_k^\dag = \lvert d_i \rangle \langle d_k \rvert$ with $\langle d_1 \rvert \propto \left( 1, 1, 0 \right)$, $\langle d_2 \rvert \propto \left( 0, 1, 1 \right)$, $\langle d_3 \rvert \propto \left( 1, 0, 1 \right)$ and $U_k$ defined in Eq.~(\ref{eq:quantonWWDinteraction}).
  The POVMs are chosen as:
  \textbf{(a)} $\Pi_1 = \cos^2 \theta \lvert d_3 \rangle \langle d_3 \rvert$,  $\Pi_2 = \sin^2 \theta \lvert \psi_{123} \rangle \langle \psi_{123} \rvert$ and $\Pi_3 = I - \Pi_1 - \Pi_2$ ;
  \textbf{(b)} $\Pi_1 = \text{diag} \{ \cos^2 \theta, 0, 0 \} + \sin^2 \theta \lvert d_1 \rangle \langle d_1 \rvert $, $\Pi_2 = \text{diag} \{  0, 0, \cos^2 \theta \} + \sin^2 \theta \lvert \psi_{12} \rangle \langle \psi_{12} \rvert $ and $\Pi_3 = I - \Pi_1 - \Pi_2$; and
  \textbf{(c)} $\Pi_1 = \cos^2 \theta \cdot I + \sin^2 \theta \lvert \psi_{123} \rangle \langle \psi_{123} \rvert$ and $\Pi_2 = I - \Pi_1$.
  Where, we denote $\lvert \psi_{123} \rangle \propto \left( \lvert d_1 \rangle + \lvert d_2 \rangle + \lvert d_3 \rangle \right)$ and $\lvert  \psi_{12}  \rangle \propto \left(   \lvert d_2 \rangle - 0.5 \lvert d_1 \rangle  \right)$.
  Before the POVM measurements, the three quantities are given as $\left[\PP \left(\rho_\text{s1}  \right) \right]^2 \approx 0.22$, $\left[\CC \left(\rho_\text{s1} \right) \right]^2 \approx 0.15$ and $S_L \left(\rho_\text{s1} \right) \approx 0.56$.
  The results show that one can change the values of $\PPbar$, $\CCbar$ and $\bar{S}_L$ by modifying POVMs.
  In particular, as shown in (b) and (c), one may keep $\PPbar$ and $\CCbar$ unchanged, see the detailed explanations in the Supplementary Material.
  }
\end{figure*}

The WWD-quanton interaction is described by
\begin{equation}
\label{eq:quantonWWDinteraction}
\hat{U}_{\text{WWD}} \equiv \sum_k \left\vert k \right> \left< k \right\vert \otimes \hat{U}_k,
\end{equation}
with the eigenstates $\lbrace \left\vert k \right> \rbrace$ for the observable $\hat{O}$ and the unitary operators $\lbrace \hat{U}_k \rbrace$ on the WWD.
This interaction will entangle the quanton and the WWD.
The input state $\rho_\text{s} \otimes \rho_\text{d}$ will evolve according to
$$\rho_\text{sd} = U_\text{WWD} \biggl(\rho_\text{s} \otimes \rho_\text{d} \biggr) U_\text{WWD}^\dag.$$
After the WWD-quanton interaction, the quanton and the WWD are respectively described by the reduced density matrices $\rho_\text{s1} = \text{tr}_\text{d} \bigl\{ \rho_\text{sd} \bigr\}$ and $\rho_\text{d1} = \text{tr}_\text{s} \bigl\{ \rho_\text{sd} \bigr\}$.
We then implement a POVM measurement $\{ \Pi_i \vert i = 1, 2, \dots, M \}$ on the WWD and sort the quanton's states into $M$ subensembles according to the applied POVM measurement.
For the $i$-th subensemble corresponding to $\Pi_i$, the quanton's reduced density matrix reads as
$$\rho_{\text{s},i} = \frac{  \text{tr}_\text{d} \{ \Pi_i \rho_{\text{sd}} \}  } {\mathbb{P} \left( \Pi_i \right) }.$$
Here $\mathbb{P} \left(\Pi_i \right) = \text{tr} \{\Pi_i \rho_\text{sd}\}$ is the probability of getting the $i$-th result corresponding to $\Pi_i$.
Therefore, averaging over all subensembles, one can obtain the average which-path predictability $\PPbar$, the average quantum coherence $\CCbar$ and the average linear entropy $\bar{ S }_L$,
\begin{subequations}
\begin{eqnarray}
\label{eq:def_avgP}
\PPbar &\equiv& \sum_{i=1}^M \left. \mathbb{P} \left( \Pi_i \right) \PP \left( \rho_{\text{s},i} \right) \right.,
\\
\label{eq:def_avgC}
\CCbar &\equiv& \sum_{i=1}^M \left. \mathbb{P} \left( \Pi_i \right) \CC \left( \rho_{\text{s},i} \right) \right.,
\\
\label{eq:def_avgS}
\bar{ S }_L &\equiv& \sum_{i=1}^M \left. \mathbb{P} \left( \Pi_i \right) S_L \left( \rho_{\text{s},i} \right) \right..
\end{eqnarray}
\end{subequations}
It's easy to prove that these three average quantities satisfy the following TCR,
\begin{equation}
\label{eq:avgPCStriality}
\left( \PPbar \right)^2   +   \left( \CCbar \right)^2   +   \bar{S}_L  \leq  1.
\end{equation}

\begin{proof}
To prove it, we should notice that the TCR~(\ref{eq:PCStriality}) for the case of no WWD is valid for arbitrary density matrices. Thus for each reduced density matrix $\rho_{\text{s},i}$, we have
$$ \mathcal{P}_i^2 + \mathcal{C}_i^2  \leq 1 - S_i $$
with $\mathcal{P}_i \equiv \PP(\rho_{\text{s},i})$, $\mathcal{C}_i \equiv \CC(\rho_{\text{s},i})$ and $S_i = S_L(\rho_{\text{s},i})$.
From $\left( \mathcal{P}_i - \mathcal{P}_k \right) ^2 \geq 0$ and $\left( \mathcal{C}_i - \mathcal{C}_k \right) ^2 \geq 0$, we have
$$\mathcal{P}_i \mathcal{P}_k + \mathcal{C}_i \mathcal{C}_k \leq \frac 12 \left(  \mathcal{P}_i^2 + \mathcal{P}_k^2 + \mathcal{C}_i^2 + \mathcal{C}_k^2 \right) \leq 1 - \frac 12 \left( S_i + S_k \right),$$
\begin{equation*}\begin{split}
\left( \PPbar \right)^2   +   \left( \CCbar \right)^2 %
&= \sum_{i,k} \mathbb{P}(\Pi_i) \mathbb{P}(\Pi_k) \left( \mathcal{P}_i \mathcal{P}_k + \mathcal{C}_i \mathcal{C}_k \right),
\\
&\leq \sum_{i,k} \mathbb{P}(\Pi_i) \mathbb{P}(\Pi_k) \left[ 1 - \frac 12 \left( S_i + S_k \right) \right],
\\
&=1 - \bar{S}_L \qedhere.
\end{split}\end{equation*}
\end{proof}

Comparing the TCRs (\ref{eq:PCStriality}) and (\ref{eq:avgPCStriality}), one may notice that the way we average the linear entropy is different from the way we average the which-path predictability and the quantum coherence.
This is because that the linear entropy has a more direct physical meaning compared to its square root.
Thus, it's more reasonable to directly sum up the linear entropy than to sum up its square root. Actually, the TCR~(\ref{eq:avgPCStriality}) is valid even if we average the linear entropy in the same way as we average the which-path predictability and the quantum coherence, that is, $\bar{S}_L$ is defined as the square of the average $\sqrt{S_i}$.
The proof is very similar to the one we give above (see details in the Supplemental Material).

Now, let's discuss the effects of the POVM measurements on WWD.
As $S_L\left( \rho \right)$ is a concave function of $\rho$, the average linear entropy is upper bounded by: $\bar{S}_L \leq S_L \left( \rho_{\text{s1}} \right)$.
Actually, since $\left. \rho_{\text{s1}} = \sum_i \mathbb{P}(\Pi_i) \rho_{\text{s},i}  \right.$, by using Eq.~(\ref{eq:concaveRelation}), we have
$$S_L(\rho_\text{s1}) = S_L\left( \sum_i \mathbb{P}(\Pi_i) \rho_{\text{s},i} \right) \geq \sum_i \mathbb{P}(\Pi_i) S_L \left( \rho_{\text{s},i} \right) = \bar{S}_L.$$
Similarly, due to $\PP (\rho)$ and $\CC (\rho)$ are convex functions of $\rho$, the which-path predictability and the quantum coherence are respectively lower bounded by: $\left. \PPbar \geq \PP (\rho_{\text{s1}}) \right.$ and $\left. \CCbar \geq \CC (\rho_{\text{s1}}) \right.$.
Thus, at the cost of decreasing the linear entropy, the POVM measurement on WWD can increase the which-path predictability and the quantum coherence.
That's to say, the POVM measurement on WWD can transform the linear entropy into the which-path predictability and the quantum coherence.
By designing the POVM measurement, one can control how much the which-path predictability and the quantum coherence to be acquired and how much the linear entropy to be sacrificed.
In Fig.~\ref{fig:POVMsweep}, we show how the POVM measurement on WWD affects the transformation among the quantum coherence, the which-path predictability and the linear entropy.
In particular, under some specific POVM measurements on WWD, one of the three quantities keeps unchanged and the wave-particle-mixedness complementarity is simplified as the particle-mixedness duality in Fig.~\ref{fig:POVMsweep}(b) or the wave-mixedness duality in Fig.~\ref{fig:POVMsweep}(c).

\emph{Connections to Other Complementarity Relations.}---
Firstly, our TCR~(\ref{eq:avgPCStriality}) naturally enclose the duality relation~(\ref{eq:EGYieq_multi}) in Ref.~\cite{Bagan16}.
From our TCR~(\ref{eq:avgPCStriality}), using the relations $\bar{S}_L \geq 0$ and
$\left. \CCbar \geq \CC (\rho_{\text{s1}}) \right.$, one can derive the following duality relation
\begin{equation}
\label{eq:avgP_Cduality}
\left( \PPbar \right) ^2 + \left[ \CC (\rho_\text{s1}) \right] ^2 \leq 1.
\end{equation}
This inequality is almost the same as the duality relation~(\ref{eq:EGYieq_multi}) in Ref.~\cite{Bagan16}.
The only difference is that Ref.~\cite{Bagan16} made an assumption on the form of the POVM measurement on WWD: They identify a click in detector $i$ with the detection of the quanton going through path $i$.
This means that the number of possible outcomes of POVM is assumed be $M=N$ and the most probable path in the $i$-th subensemble is the $i$-th path.
Under this assumption, our TCR~(\ref{eq:avgP_Cduality}) agrees with the duality relation~(\ref{eq:EGYieq_multi}) in Ref.~\cite{Bagan16}.
Differently, in Ref.~\cite{Bagan16}, the duality relation~(\ref{eq:EGYieq_multi}) is proven by introducing an loose upper bound of $\PPbar$ in the sense of minimum-error state discrimination.

Secondly, our wave-particle-mixedness complementarity relates to the wave-particle-entanglement complementarity for bi-partite systems~\cite{Oppenheim03,Peng05,Melo07,Peng08,Jakob10,Fedrizzi11}.
Comparing our TCR~(\ref{eq:PCStriality}) with the the wave-particle-entanglement complementarity relation~\cite{Melo07,Jakob10}, our density matrix corresponds to the single-partite reduced density matrix for the bi-partite systems.
Therefore, the mixedness aspect described by linear entropy corresponds to the bi-partite entanglement described by concurrence.
Considering two entangled qubits and choosing concurrence as the measure of bi-partite entanglement, the wave-particle-entanglement complementarity relation has been given as~\cite{Melo07,Jakob10}
\begin{equation}
\label{eq:concurrenceTriality}
\left(\PP \right)^2 + \left(\CC \right)^2 + \mathscr{C}^2 \leq 1.
\end{equation}
Here, $\PP$ and $\CC$ are respectively the which-path predictability and the quantum coherence for the single-partite reduced density matrix, and $\mathscr{C}$ is the concurrence between the two entangled qubits.
If the two-qubit system is in a pure state, we exactly have the linear entropy $S_L(\rho_i) = \mathscr{C}^2$ with $\rho_i$ denoting the reduced density matrix for one of the two qubits.
This means that our wave-particle-mixedness complementarity relation~(\ref{eq:PCStriality}) coincides with the wave-particle-entanglement complementarity relation~(\ref{eq:concurrenceTriality}) for two-qubit pure states.
When the which-path predictability is fixed, our TCR gives a wave-mixedness duality relation, which corresponds to the complementarity between entanglement and coherence~\cite{Singh15,Streltsov16,Chitambar16}.
However, when the bi-partite system is in a mixed state, these two complementarity relations become different: our TCR~(\ref{eq:PCStriality}) is still an equality while the wave-particle-entanglement complementarity relation becomes an inequality.
This indicates that, for a two-qubit system, we always have $S_L(\rho_i) \geq \mathscr{C}^2$.

\emph{Conclusions.}---In conclusion, for an arbitrary multi-state quanton, we have explored the wave-particle-mixedness complementarity by analyzing the quantum coherence, the which-path predictability and the linear entropy.
Beyond the conventional wave-particle dulaity, our results indicate that wave, particle and mixedness aspects are three complementary properties.
By introducing the WWD-quanton interaction, one can control the values of these three quantities by modifying the POVM on WWD.
Through designing proper POVM measurements on WWD, one can transform the linear entropy into the which-path predictability or the quantum coherence.
Moreover, our wave-particle-mixedness complementarity relations naturally enclose the generalized wave-particle duality relation for multi-state systems~\cite{Bagan16} and are closely related to the wave-particle-entanglement complementarity relation for bi-partite systems~\cite{Oppenheim03,Peng05,Melo07,Peng08,Jakob10,Fedrizzi11}.

\begin{acknowledgments}
This work was supported by the National Natural Science Foundation of China (Grants No. 11374375 and No. 11574405).
J. H. is partially supported by National Postdoctoral Program for Innovative Talents of China (Grant No. BX201600198).
\end{acknowledgments}


%
\appendix

\bigskip

\section*{Supplementary Material}

\renewcommand{\thefigure}{S\arabic{figure}}
 \setcounter{figure}{0}
\renewcommand{\theequation}{S.\arabic{equation}}
 \setcounter{equation}{0}
 \renewcommand{\thesection}{S.\Roman{section}}
\setcounter{section}{0}
\setcounter{lemma}{0}

\section{\label{app:lemmaProof}Appendix A: Proof of two lemmas}
In this appendix, we will prove the two lemmas mentioned in Section II 
 of the main text.
\begin{lemma}
For an arbitrary $N \times N$ density matrix $\rho$, we always have
\begin{equation}
\label{eq:lemmaDiag}
\sum_{k=1}^{N} \left( \rho _{kk} \right) ^2 \geq p_1^2 + \frac{\left( 1 - p_1 \right) ^2}{N-1}
\end{equation}
with $p_1 = \max_k \rho _{kk}$ being the largest diagonal element of $\rho$.
The equal sign holds if and only if all diagonal elements of $\rho$, except the maximal one, have the same value.
\begin{proof}
For simplicity, let's assume $\rho_{11}$ to be the largest diagonal element, i.e. $\rho_{11} = p_1$. Then we have $\sum_{k=2}^N \rho_{kk} = 1-p_1$. Using Cauchy-Schwarz inequality:
$$ \left(  \sum_{k=2}^N \left( \rho_{kk} \right)^2  \right) \left(   \sum_{k=2}^N 1^2   \right) \geq \left(   \sum_{k=2}^N \rho_{kk} \cdot 1  \right) ^2 = \left( 1 - p_1 \right)^2 $$
It immediately follows that
$$ \sum_{k=1}^N \left( \rho_{kk} \right)^2  =  p_1^2 + \sum_{k=2}^N \left( \rho_{kk} \right)^2 \geq p_1^2 + \frac{\left( 1-p_1 \right)^2}{N-1} $$
The equal sign holds if and only if $\rho_{22} = \rho_{33} = \dots = \rho_{NN} = \left. \left( 1 - p_1 \right) / \left( N - 1 \right) \right.$.
\end{proof}
\end{lemma}
\begin{lemma}
For an arbitrary $N \times N$ density matrix $\rho$, we always have
\begin{equation}
\label{eq:lemmaOffdiag}
\sum _{i \neq k} \sum _{l \neq m} \lvert \rho _{ik} \rvert \cdot \lvert \rho _{lm} \rvert \leq N(N-1)\sum _{i \neq k} \lvert \rho _{ik} \rvert ^2.
\end{equation}
The equal sign holds if and only if all off-diagonal elements of $\rho$ have the same modulus.
\begin{proof}
For simplicity, let's denote the set $\{ \lvert \rho_{ik} \rvert, i \neq k \}$ as an $L=N\left( N-1 \right)$ dimension vector $\vec{A}$. Then Eq.~(\ref{eq:lemmaOffdiag_S}) becomes
$$ \sum_{i=1}^L \sum_{k=1}^L A_i \cdot A_k \leq N\left( N-1 \right) \lvert \vec{A} \rvert ^2$$
The LHS of the above inequality is the summation of all the matrix elements of the following $L \times L$ matrix:
$$ \begin{pmatrix}
A_1 A_1  &  A_1 A_2  &  \dots    &  A_1 A_L    \\
A_2 A_1  &  A_2 A_2  &              &  \vdots    \\
\vdots     &                  &  \ddots  &  A_{L-1} A_L    \\
A_L A_1  &  \dots       & A_L A_{L-1}  &  A_L A_L
\end{pmatrix} $$
The summation can be divided into $L$ parts:
\begin{equation}
\label{eq:sumArray_S}
\begin{array}{lllllllllll}
A_1 A_1 &+& A_2 A_2 &+& \dots &+& A_{L-2} A_{L-2} &+& A_{L-1} A_{L-1} &+& A_L A_L \\
A_2 A_1 &+& A_3 A_2 &+& \dots &+& A_{L-1} A_{L-2} &+& A_L A_{L-1} &+& A_1 A_L \\
A_3 A_1 &+& A_4 A_2 &+& \dots &+& A_L A_{L-2} &+& A_1 A_{L-1} &+& A_2 A_L \\
               &   &              & &\vdots &  &                      & &                       & &      \\
A_L A_1 &+& A_1 A_2 &+& \dots &+& A_{L-3} A_{L-2} &+& A_{L-2} A_{L-1} &+& A_{L-1} A_L \\
\end{array}
\end{equation}
By Cauchy-Schwarz inequality, we can see that each row in Eq.~(\ref{eq:sumArray_S}) has an upper bound of $\lvert \vec{A} \rvert^2$. For example, the third line:
\begin{equation*}
\begin{split}
A_3 A_1 +& A_4 A_2 + \dots+ A_L A_{L-2} + A_1 A_{L-1} + A_2 A_L \\
          \leq& \sqrt{A_3^2+A_4^2+\dots+A_L^2+A_1^2+A_2^2} \sqrt{A_1^2+A_2^2+\dots+A_L^2} \\
            =& \lvert \vec{A} \rvert ^2
\end{split}
\end{equation*}
So we have
$$ \sum_{i=1}^L \sum_{k=1}^L A_i \cdot A_k \leq L \lvert \vec{A} \rvert ^2 = N(N-1) \lvert \vec{A} \rvert ^2 $$
The equal sign holds if and only if all the summation in Eq.~(\ref{eq:sumArray_S}) reach their upper bound, i.e.
$$ A_1 = A_2 = \dots = A_L $$
That is: all the off-diagonal elements of $\rho$ have the same modulus.
\end{proof}
\end{lemma}

\section{\label{app:concaveProof}Appendix B: Concave property of linear entropy}
In this appendix, we will prove that $S_L(\rho)$ is a concave function of $\rho$:
\begin{equation}
\label{eq:concaveRelation_S}
\begin{split}
S_L\left( a \rho_1 + b \rho_2 \right) \geq  a S_L \left( \rho_1 \right) &+ b S_L \left( \rho_2 \right)
\\
&a + b = 1, a \in \left[ 0, 1 \right]
\end{split}
\end{equation}
\begin{proof}
Note that $\text{tr} \{ \rho^2 \} = \sum_{i,k} \lvert \rho_{ik} \lvert ^2$ and $a+b=1$, Eq.~(\ref{eq:concaveRelation_S}) can expand to the following form:
\begin{equation}
\label{eq:concaveRelation2_S}
\sum_{i,k} \lvert a \left( \rho_1 \right) _{ik} + b \left( \rho_2 \right) _{ik} \lvert ^2   \leq   %
a \left. \sum_{i,k} \lvert \left( \rho_1 \right) _{ik} \rvert ^2  \right.  + %
b \left. \sum_{i,k} \lvert \left( \rho_2 \right) _{ik} \rvert ^2  \right.
\end{equation}
Our task then becomes to prove Eq.~(\ref{eq:concaveRelation2_S}). The difference between the LHS and RHS of Eq.~(\ref{eq:concaveRelation2_S}) is shown as Eq.~(\ref{eq:diffConcaveR2_S}).
\begin{widetext}
\begin{subequations}
\label{eq:diffConcaveR2_S}
\begin{align}
\text{LHS - RHS} &= a (a-1) \left. \sum_{i,k} \lvert \left( \rho_1 \right) _{ik} \rvert ^2  \right.  +%
    b (b-1) \left. \sum_{i,k} \lvert \left( \rho_2 \right) _{ik} \rvert ^2  \right.  +%
    2 a b \left. \sum_{i,k} \lvert \left( \rho_1 \right) _{ik} \left( \rho_2 \right) _{ik}  \rvert \right. \\
&= a (a-1) \lvert \vec{A}_1 \rvert ^2  +%
    b (b-1) \lvert \vec{A}_2 \rvert ^2  +%
    2 a b \vec{A}_1 \cdot \vec{A}_2 \\
\label{eq:dCR2_3_S}
&= a (a-1) \lvert \vec{A}_1 \rvert ^2  +%
    a (a-1) \lvert \vec{A}_2 \rvert ^2  -%
    2 a (a-1) \vec{A}_1 \cdot \vec{A}_2 \\
&= a (a-1) \left( \vec{A}_1 - \vec{A}_2 \right) ^2 \\
\label{eq:dCR2_5_S}
&\leq 0
\end{align}
\end{subequations}
\end{widetext}
Here, $\vec{A}_1 = \left(   \lvert (\rho_1)_{11} \rvert, \lvert (\rho_1)_{12} \rvert , \dots , \lvert (\rho_1)_{NN} \rvert    \right)$ is an $N^2$ dimension vector whose components are the modulus of $\rho_1$'s matrix elements (the order doesn't matter). $\vec{A}_2$ has the same relation to $\rho_2$. We get Eq.~(\ref{eq:dCR2_3_S}) by applying $a+b=1$ and the unequal sign in Eq.~(\ref{eq:dCR2_5_S}) is because $a \in [0,1]$. So we have $\text{LHS} \leq \text{RHS}$ for Eq.~(\ref{eq:concaveRelation2_S}) and the proof is finished.
\end{proof}
%
%
\begin{figure*}
 \includegraphics[width=\textwidth]{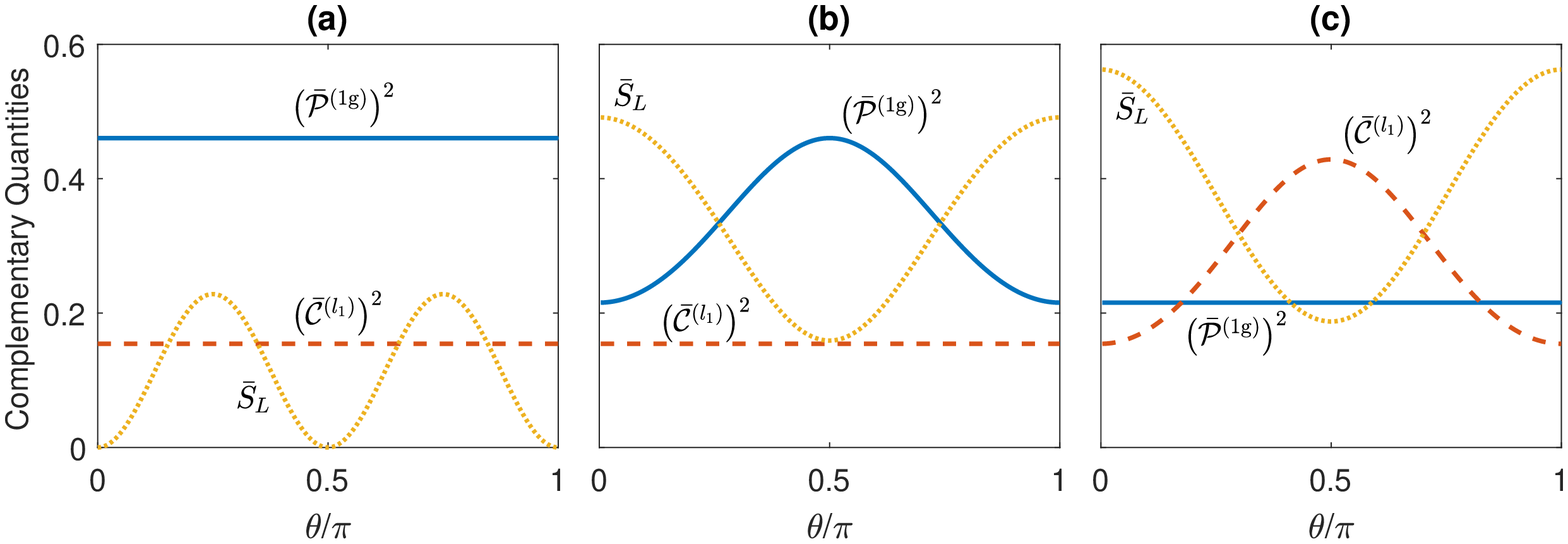}
\caption{\label{fig:POVMsweep_append}
  Another example for the change of quantum coherence, which-path predictability and linear entropy under POVM measurements on WWD.
  The quanton's initial state is $\rho_\text{s}=\lvert \psi_\text{s}\rangle \langle \psi_\text{s} \rvert$ with $\langle \psi_\text{s} \rvert \propto \left( 1, 3, 2 \right)$.
  The WWD's states are $U_i \rho_\text{d} U_k^\dag = \lvert d_i \rangle \langle d_k \rvert$ with $\langle d_1 \rvert \propto \left( 1, 1, 0 \right)$, $\langle d_2 \rvert \propto \left( 0, 1, 1 \right)$, $\langle d_3 \rvert \propto \left( 1, 0, 1 \right)$ and $U_k$ defined in Eq.~(\ref{eq:quantonWWDinteraction}).
  The POVMs are chosen as:
  \textbf{(a)} $\Pi_1 = \text{diag} \{ 1, 0, 0 \}$,
$\Pi_2 = \text{diag} \{ 0, \cos^2 \theta, \sin^2 \theta \}$ and
$\Pi_3 = \text{diag} \{ 0, \sin^2 \theta, \cos^2 \theta \}$;
  \textbf{(b)} $\Pi_1 = \text{diag} \{ \sin^2 \theta, 0, 0 \}$,
$\Pi_2 = \text{diag} \{ 0.5 \cos^2 \theta, 0, 0.5 \}$ and
$\Pi_3 = \text{diag} \{ 0.5 \cos^2 \theta, 1, 0.5 \}$;
and
  \textbf{(c)} $\Pi_1 = \cos^2 \theta \cdot I$,
$\Pi_2 = \sin^2 \theta \lvert \psi_{123} \rangle \langle \psi_{123} \rvert$ and
$\Pi_3 = \sin^2 \theta \cdot \left( I - \lvert \psi_{123} \rangle \langle \psi_{123} \rvert \right)$.
  Where, we denote $\lvert \psi_{123} \rangle \propto \left( \lvert d_1 \rangle + \lvert d_2 \rangle + \lvert d_3 \rangle \right)$.
  Before the POVM measurements, the three quantities are given as $\left[\PP \left(\rho_\text{s1}  \right) \right]^2 \approx 0.22$, $\left[\CC \left(\rho_\text{s1} \right) \right]^2 \approx 0.15$ and $S_L \left(\rho_\text{s1} \right) \approx 0.56$.
  }
\end{figure*}

\section{\label{app:boundTightness}Appendix C: Tightness of TCR (6)}
\begin{figure}[!h]
\includegraphics[width=0.5 \textwidth]{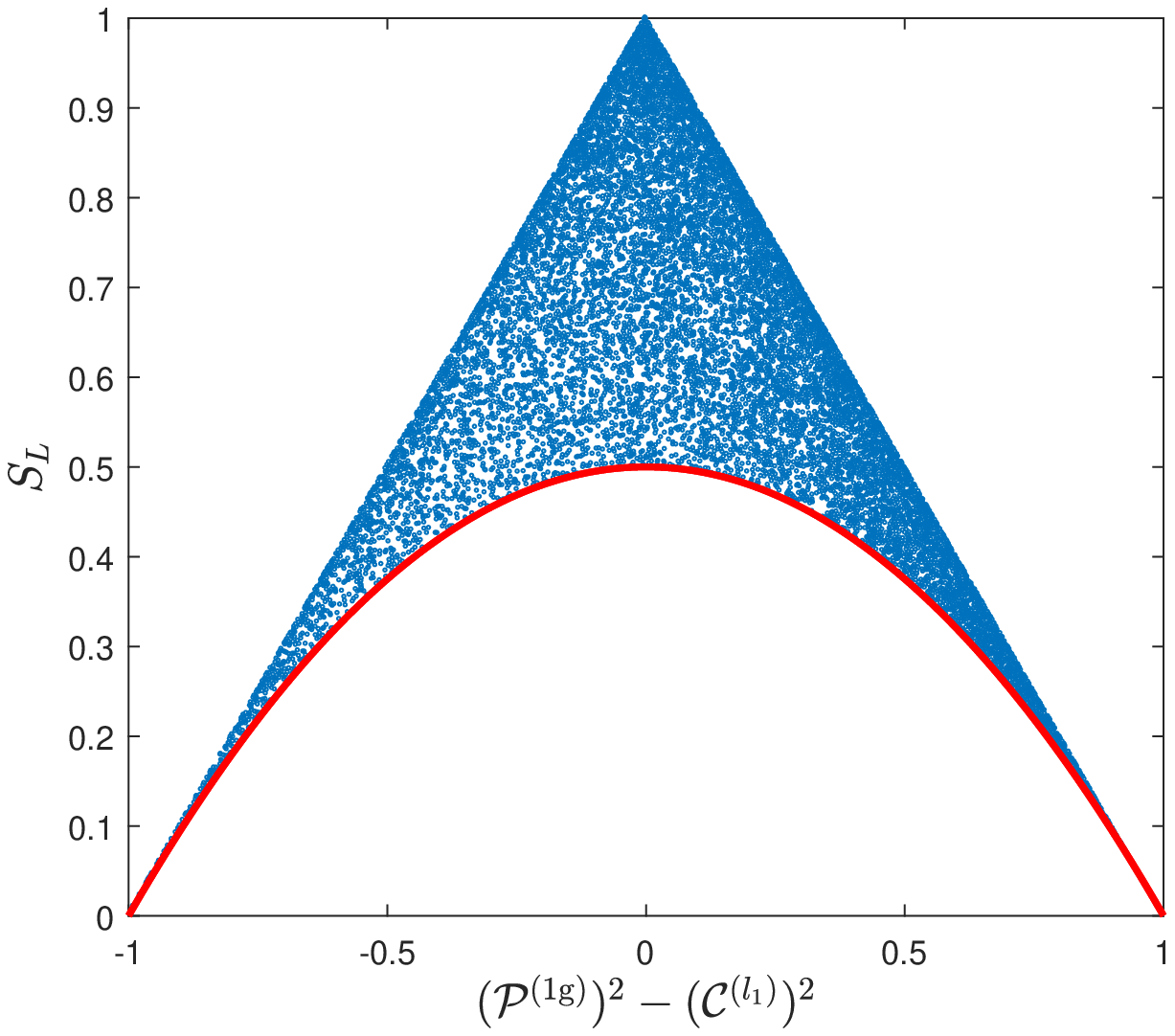}%
\caption{\label{fig:validArea}%
The reachable area on triangle $\Delta$.}
\end{figure}
\begin{equation}
\label{eq:PCStriality_S}
\left(  \PP  \right) ^2 + \left(  \CC  \right) ^2 + S_L \leq 1
\end{equation}
The TCR Eq.~(\ref{eq:PCStriality_S}) gives a bound for which-path predictability, quantum coherence and linear entropy, but is it tight? For the case of $N=2$, this bound is tight because Eq.~(\ref{eq:PCStriality_S}) becomes an equality and all states saturate it. Things get a bit complicated when $N>2$. For an ideal upper bound, we may expect that all points on the triangle
$$\Delta= \bigg\lbrace (\PP, \CC, S_L) \in [0,1]^3 \bigg\vert %
     \left( \PP \right) ^2 +  \left( \CC \right) ^2 + S_L = 1  \bigg\rbrace $$
represent actually reachable combination of $\PP$, $\CC$ and $S_L$. We visualize this trialgle in Fig.~\ref{fig:validArea}. Each blue point in Fig.~\ref{fig:validArea} represents one density matrix that has the form of Eq.~(\ref{eq:eqsignHold}) in main text. We randomly generate many density matrices of this form and found that there is an area where no density matrix can reach. The boundary of this area is highlighted by red line in Fig.~{\ref{fig:validArea}}. Its analytic form is
$$2 S_L = 1- \left[  \left( \PP \right)^2 - \left( \CC \right)^2 \right] ^2$$
This is not hard to verify. By calculating the $\PP$, $\CC$ and $S_L$ of Eq.~(\ref{eq:eqsignHold}), we found that
\begin{align*}
\left[   \left( \PP \right)^2   - \left( \CC \right)^2    \right] ^2 - \left( 1-2 S_L \right) &= (1-x)^2 + (1-y)^2 - 2 x y - 1 \\
x = \left( N p_2 - 1 \right)^2,  &\qquad y = N^2 \lvert a \rvert ^2
\end{align*}
Then by applying the positive-semidefinite condition $\lvert a \rvert \leq p_2$, we have:
\begin{align*}
1 - y &\geq 1 - N^2 p_2^2 \geq 0 \\
- 2 x y &\geq - 2 x N^2 p_2^2 \\
\left[   \left( \PP \right)^2   - \left( \CC \right)^2    \right] ^2 -& \left( 1-2 S_L \right) \\
&\geq (1-x)^2 + (1 - N^2 p_2^2 )^2 - 2 x N^2 p_2^2 - 1 \\
&=0
\end{align*}
So for all density matrices that satisfy Eq.~(\ref{eq:eqsignHold}),
$$2 S_L \geq 1 - \left[ \left( \PP \right)^2 - \left( \CC \right)^2 \right]^2$$

The above result shows that the upper bound in Eq.~(\ref{eq:PCStriality_S}) is tight only when $S_L$ is larger than 0.5. When $S_L < 0.5$, some parts of the bound are actually not reachable.

\section{\label{app:avgNotChange}Appendix D: Explanation of constant $\CCbar$ and $\PPbar$ in Fig.~\ref{fig:POVMsweep}}
Consider a quanton with 3 DOFs, and  the WWD has the property $U_i \rho_\text{d} U_k^\dag = \lvert d_i \rangle \langle d_k \rvert$. Then we have
\begin{equation}
\label{eq:rho_sdForm1}
\rho_{\mathrm{sd}}=
\begin{pmatrix}
       \rho_{11} \left| d_1\right\rangle \left\langle d_1\right| & \rho_{12} \left| d_1\right\rangle \left\langle d_2\right|  & \rho_{13} \left| d_1\right\rangle \left\langle d_3\right|  \\
       \rho_{21} \left| d_2\right\rangle \left\langle d_1\right| & \rho_{22} \left| d_2\right\rangle \left\langle d_2\right|  & \rho_{23} \left| d_2\right\rangle \left\langle d_3\right|  \\
       \rho_{31} \left| d_3\right\rangle \left\langle d_1\right| & \rho_{32} \left| d_3\right\rangle \left\langle d_2\right|  & \rho_{33} \left| d_3\right\rangle \left\langle d_3\right|
\end{pmatrix}
\end{equation}
\begin{equation}
\nonumber
\rho_{\mathrm{s},i}=\frac{  \mathrm{tr_d}\{   \Pi_i \rho_\mathrm{sd}  \} } {\mathbb{P}(\Pi_i)}, \quad i=1,2,3
\end{equation}

\paragraph{For $\PPbar $}
\begin{equation}
\nonumber
\PPbar \equiv \sum_{i=1}^3 \left. \mathbb{P} \left( \Pi_i \right) \PP \left( \rho_{\text{s},i} \right) \right.
= \frac{3}{2} \sum_{i=1}^3 \max_k{ \left(       \mathrm{tr_d}\{   \Pi_i \rho_\mathrm{sd}  \}      \right) _ {kk} }   - \frac{1}{2}
\end{equation}
If the positions of maximal diagonal element in all subensembles are the same, or more accuately, if
\begin{equation}
\nonumber
\max_k{ \left(       \mathrm{tr_d}\{   \Pi_i \rho_\mathrm{sd}  \}      \right) _ {kk} }  =   \left(       \mathrm{tr_d}\{   \Pi_i \rho_\mathrm{sd}  \}      \right) _ {k_0 k_0},  \forall i
\end{equation}
$k_0$ does not depend on $i$. Then we have
\begin{equation}
\label{eq:PPbarUnchange}
\begin{split}
\PPbar =& \frac{3}{2} \sum_i {  \left(       \mathrm{tr_d}\{   \Pi_i \rho_\mathrm{sd}  \}      \right) _ {k_0 k_0} }  -\frac{1}{2}  %
\\
=& \frac{3}{2}  \left(       \mathrm{tr_d}\{ \rho_\mathrm{sd}  \}      \right) _ {k_0 k_0} - \frac{1}{2} %
= \PP \left(  \rho_\mathrm{s1} \right)
\end{split}
\end{equation}
which is independent of the choice of POVM measurement $ \left\{  \Pi_i \right\} $. This means if the measurement does not perform strong detection on path information, i.e. does not change the most probable path, we may expect $\PPbar$ to be unchanged. The last equal sign above is because the diagonal elements of $   \mathrm{tr_d}\{   \Pi_i \rho_\mathrm{sd}  \}   $ are always positive, so $  \left(   \mathrm{tr_d}\{ \rho_\mathrm{sd}  \}  \right) _ {k_0 k_0}  $ is the maximal diagonal element of $ \rho_\mathrm{s1} $.

\paragraph{For $\CCbar$}
\begin{equation}
\nonumber
\CCbar=\sum_i { \mathbb{P}(\Pi_i) \CC(\rho_{\mathrm{s},i}) }  %
=\frac{1}{N-1} \sum_{j \neq k} { \sum_{i=1}^3{  \left| \left(  \mathrm{tr_d}\left\{  \Pi_i \rho_\mathrm{sd}  \right\} \right) _{jk}  \right|  } }
\end{equation}
If $\left(  \mathrm{tr_d}\left\{  \Pi_1 \rho_\mathrm{sd}  \right\} \right) _{jk},  %
      \left(  \mathrm{tr_d}\left\{  \Pi_2 \rho_\mathrm{sd}  \right\} \right) _{jk}  \textrm{and}  %
      \left(  \mathrm{tr_d}\left\{  \Pi_3 \rho_\mathrm{sd}  \right\} \right) _{jk} \left( j \neq k \right) $ have the same phase angle, then we have
\begin{equation}
\label{eq:CCbarUnchange}
\begin{split}
\CCbar=&\frac{1}{N-1} \sum_{j \neq k} { \left|  \sum_{i=1}^3{  \left(  \mathrm{tr_d}\left\{  \Pi_i \rho_\mathrm{sd}  \right\} \right) _{jk} } \right|   }  %
\\
=&\frac{1}{N-1} \sum_{j \neq k} { \left| \left(  \mathrm{tr_d}\left\{  \rho_\mathrm{sd}  \right\} \right) _{jk}  \right|   }   %
=\CC (  \rho_\mathrm{s1} )
\end{split}
\end{equation}
which is also independent of the choice of POVM measurement.

Note that
$\left(  \mathrm{tr_d}\left\{  \Pi_i \rho_\mathrm{sd}  \right\} \right) _{jk} = \mathbb{P}(\Pi_i) \left( \rho_{\mathrm{s},i} \right)_{jk}$
are the off-diagonal elements of the subensembles' reduced density matrix. So a physical understanding of the above criterion is that if the measurement on WWD do not change the phase of the quanton's density matrix, then $\CCbar$ will keep constant.

Using Eq.~(\ref{eq:rho_sdForm1}), we get
\begin{equation}
\nonumber
\left(  \mathrm{tr_d}\left\{  \Pi_i \rho_\mathrm{sd}  \right\} \right) _{jk} %
= \rho_{jk} \left\langle d_k \middle\vert \Pi_i \middle\vert d_j  \right\rangle
\end{equation}
Moreover, if we choose $ \left\{  \Pi_i  \right\} $ to be projectors: $\Pi_i=\left\vert \pi_i \middle\rangle \middle \langle \pi_i \right\vert $ Then
\begin{equation}
\nonumber
\left(  \mathrm{tr_d}\left\{  \Pi_i \rho_\mathrm{sd}  \right\} \right) _{jk} %
=   \rho_{jk} \left\langle \pi_i  \middle\vert d_j  \middle\rangle \middle\langle d_k \middle\vert \pi_i \right\rangle
\end{equation}
So the criterion for Eq.~(\ref{eq:CCbarUnchange}) is: the three diagonal elements of matrix $\left\vert d_j  \middle\rangle \middle\langle d_k \right\vert $ under basis $\left\{  \left\vert \pi_i \right\rangle \right\}$ have the same phase angle. This can be used to construct $\{\Pi_i\}$

\paragraph{About how to plot Fig.~(\ref{fig:POVMsweep}) }
The problem is to find a continuum of POVM that makes Eq.~(\ref{eq:PPbarUnchange}) or Eq.~(\ref{eq:CCbarUnchange}) valid. Suppose we find two set of POVM $ \left\{ \Pi_i \right\} $ and $ \left\{ \Lambda_i \right\}  $ which make Eq.~(\ref{eq:PPbarUnchange}) valid and the $k_0$ for them is the same, then it's easy to check that a linear combination of $\Pi_i$ and $\Lambda_i$ also makes Eq.~(\ref{eq:PPbarUnchange}) valid. So we can just choose the POVM $ \left\{ \Pi_i \cos^2{\theta}  + \Lambda_i \sin^2{\theta}  \right\} $ and vary $\theta$.

 Similarly, if we find two set of POVM $ \left\{ \Pi_i \right\} $ and $ \left\{ \Lambda_i \right\}  $ which make Eq.~(\ref{eq:CCbarUnchange}) valid and the phase angle of $\left(  \mathrm{tr_d}\left\{  \Pi_i \rho_\mathrm{sd}  \right\} \right) _{jk}$ and $\left(  \mathrm{tr_d}\left\{  \Lambda_i \rho_\mathrm{sd}  \right\} \right) _{jk}$ are the same, then again $ \left\{ \Pi_i \cos^2{\theta}  + \Lambda_i \sin^2{\theta}  \right\} $ is what we need. More examples are shown in Fig.~(\ref{fig:POVMsweep_append}).

\section{\label{app:anotherTCR}Appendix E: Another TCR from a different averaging method for linear entropy}
In the main text, we mentioned that it's possible to get a similar TCR if we define $\bar{S}_L$ as the square of the average $\sqrt{S_i}$, i.e.
$$ \bar{ S }'_L \equiv \sum_{i=1}^M \left. \mathbb{P} \left( \Pi_i \right) \sqrt{ S_L \left( \rho_{\text{s},i} \right) } \right. $$
The new TCR is
\begin{equation}
\label{eq:avgPCStriality_new}
\left( \PPbar \right)^2 + \left( \CCbar \right)^2 + \bar{S}'_L \leq 1
\end{equation}
\begin{proof}
The proof is very similar to the one in main text. Denote $\mathcal{P}_i \equiv \PP(\rho_{\text{s},i})$, $\mathcal{C}_i \equiv \CC(\rho_{\text{s},i})$ and $\mathcal{E}_i \equiv \sqrt{ S_L(\rho_{\text{s},i}) }$. Since Eq.~(\ref{eq:PCStriality}) is valid for arbitrary density matrix, for each subensemble we have
$$ \mathcal{P}_i^2 + \mathcal{C}_i^2 + \mathcal{E}_i^2  \leq 1$$
And from the fact that $\left( \mathcal{P}_i - \mathcal{P}_k \right) ^2 \geq 0$, $\left( \mathcal{C}_i - \mathcal{C}_k \right) ^2 \geq 0$ and $\left( \mathcal{E}_i - \mathcal{E}_k \right) ^2 \geq 0$, we have
$$ \mathcal{P}_i \mathcal{P}_k + \mathcal{C}_i \mathcal{C}_k + \mathcal{E}_i \mathcal{E}_k \leq \frac 12 \left(  \mathcal{P}_i^2 + \mathcal{P}_k^2 + \mathcal{C}_i^2 + \mathcal{C}_k^2 + \mathcal{E}_i^2 + \mathcal{E}_k^2 \right) \leq 1$$
\begin{equation*}\begin{split}
\left( \PPbar \right)^2   +   \left( \CCbar \right)^2 + \bar{S}'_L %
&= \sum_{i,k} \mathbb{P}(\Pi_i) \mathbb{P}(\Pi_k) \left( \mathcal{P}_i \mathcal{P}_k + \mathcal{C}_i \mathcal{C}_k + \mathcal{E}_i \mathcal{E}_k \right)
\\
&\leq \sum_{i,k} \mathbb{P}(\Pi_i) \mathbb{P}(\Pi_k)
\\
&=1  \qedhere
\end{split}\end{equation*}
\end{proof}
At last, we also want to mention that $\mathcal{E}\left( \rho \right) \equiv \sqrt{ S_L \left( \rho \right) }$ is a concave function of $\rho$:
\begin{equation}
\label{eq:concaveRelation2}
\begin{split}
\mathcal{E}\left( a \rho_1 + b \rho_2 \right) \geq  a \mathcal{E} \left( \rho_1 \right) &+ b \mathcal{E} \left( \rho_2 \right)
\\
&a + b = 1, a \in \left[ 0, 1 \right]
\end{split}
\end{equation}
So the discussion about the effects of the POVM measurement on WWD in the main text also applied to this new TCR Eq.~(\ref{eq:avgPCStriality_new}).

\end{document}